\documentstyle[epsfig,11pt]{article}
\newcommand{\beq}{\begin{equation}}
\newcommand{\eeq}{\end{equation}}
\newcommand{\ben}{\begin{eqnarray}}
\newcommand{\een}{\end{eqnarray}}
\newcommand{\bc}{\begin{center}}
\newcommand{\ec}{\end{center}}

\begin{document}

\title{Self-existing objects and auto-generated
information in chronology-violating space-times: A philosophical
discussion}
\author{Gustavo E. Romero\thanks{Member of CONICET. E-mail:
romero@irma.iar.unlp.edu.ar}  and Diego F. Torres\thanks{E-mail:
dtorres@venus.fisica.unlp.edu.ar}}
\maketitle \vspace{-1cm}

\begin{center}
{\sl Instituto Argentino de Radioastronom\'{\i}a, C.C.5, (1894)
Villa Elisa, Buenos Aires, Argentina}\end{center}

\begin{abstract}

Closed time-like curves naturally appear in a variety of
chronology-violating space-times. In these space-times, the
Principle of Self-Consistency demands an harmony between local and
global affairs that excludes grandfather-like paradoxes. However,
self-existing objects trapped in CTCs are not seemingly avoided by
the standard interpretation of this principle, usually constrained
to a dynamical framework. In this paper we discuss whether we are
committed to accept an ontology with self-existing objects if CTCs
actually occur in the universe. In addition, the epistemological
status of the Principle of Self-Consistency is analyzed and a
discussion on the information flux through CTCs is presented.
\end{abstract}

\section{Introduction}

One of the most fascinating aspects of the general theory of
relativity (GTR) is that its field equations have solutions where
the formation of closed time-like curves (CTCs) is possible. These
curves represent the world lines of any physical system in a
temporally orientable space-time that, moving always in the future
direction, ends arriving back at some point of its own past.
Although solutions of the Einstein field equations (EFEs) where
CTCs exist are known at least since G\"odel's (1949) original
work, it has been just in the last dozen of years that physicists
have shown  a strong and sustained interest on this topic. The
renewed attraction of CTCs and their physical implications stem
from the discovery, at the end of the 1980's, of traversable
wormhole space-times (Morris et al. 1988). The subject, however,
have called the attention of philosophers during several decades,
who have discussed paradoxes of time travel in connection with
causality and logical problems (see Nahin 1999 for an extensive
survey of the literature). Their treatment, unfortunately, has
seldom taken into account physical considerations (see,
notwithstanding, the illuminating analysis by
Earman 1995a, b, among others).\\

In the present paper, we shall discuss two problems related with
the possible existence of CTCs in GTR, which have important
implications for two of the main branches of philosophy: ontology
and epistemology. These problems can be formulated through the
following couple of questions: \begin{enumerate} \item Are we
committed, in case that CTCs actually exist in the universe, to
accept an ontology with self-existing objects? \item Can
information be generated from nothing?\end{enumerate} A
self-existing object is defined as any physical system that exists
over a finite region of space-time but has neither starting nor
ending temporal points: it is never created and it is never
destroyed, but there is at least a time at which it does not
exist. The first half of this paper will try to clarify the
meaning of the questions mentioned above. The second half will try
out some possible answers. At the end, we hope to have provided
the reader with the sense that philosophical problems arisen from
complex physical theories cannot be solved without appealing to
the technical and conceptual tools that those same theories
provide.

\section{Against CTCs: chronology protection and paradoxes}

A relativistic space-time is represented by a four-dimensional
manifold $M$ equipped with a Lorentzian metric $g_{ab}$. Any
space-time $(M,\;g_{ab})$ with CTCs is called a
chronology-violating space-time. There are two types of these
space-times: those where CTCs exist everywhere (like, for
instance, in G\"odel space-time), and those where CTCs are
confined within some regions and there exists at least one region
free of them. The regions with CTCs are separated from the
``well-behaved'' space-time by Cauchy horizons (wormhole
space-times belong to this latter type). Here we shall restrict
ourselves to the second type of space-times.\\

The existence of CTCs and the possibility of backward time travel
have been objected by several scientists championed by Hawking
(1992), who proposed the so-called chronology protection
conjecture:  the laws of physics are such that the appearance of
CTCs is never possible. The suggested mechanism to enforce
chronology protection is the back-reaction of vacuum polarization
fluctuations: when the renormalized energy-momentum tensor is fed
back to the semi-classical EFEs, the back-reaction accumulates
energy in such a way that it may distort the space-time geometry
so strongly as to form a singularity, destroying the CTC at the
very moment of its formation.\\

It has been argued, however, that quantum gravitational effects
would cut the divergence off saving the CTCs (Kim and Thorne
1991). By other hand, Li et al. (1993) pointed out that the
divergence of the energy-momentum tensor does not prevent the
formation of a CTC but is just {\sl saying us} that a full (yet
unknown) quantum gravity theory must be applied. According to
them, singularities, far from being physical entities that can act
upon surrounding objects, are manifestations of the breakdown of
the gravitational theory. In any case, we cannot draw definitive
conclusions with the semi-classical tools at our disposal (see
Earman 1995b for additional discussion).\\

But even if the energy-momentum tensor of vacuum polarization
diverges at the Cauchy horizon it is not necessarily implied that
CTCs must be destroyed, since the equations can be well-behaved in
the region inside the horizon (Li et al. 1993). In particular,
wormhole space-times could be stabilized against vacuum
fluctuations introducing reflecting boundaries between the
wormhole mouths (Li 1994) or using several wormholes to create
CTCs (Thorne 1992, Visser 1997).\\

Recently, even Hawking has recognized that back-reaction does not
necessarily enforce chronology protection (Cassidy and Hawking
1998). Although the quest for finding an effective mechanism to
avoid CTCs continues, it is probable that the definitive solution
to the problem should wait until a complete theory of quantum
gravity can be formulated. In the meantime, the profound physical
consequences of time travel in GTR should be explored in order to
push this theory to its ultimate limits, to the region where the
very foundations of the theory must be revisited.\\

A different kind of objection to CTC formation is that they allow
illogical situations like the ``grandfather'' paradox
\footnote{The grandfather paradox: A time traveler goes to his
past and kills his young grandfather then avoiding his own birth
and, consequently, the time travel in which he killed his
grandfather.} usually interpreted as the statement that the
corresponding solutions of the EFEs are ``non-physical''. This is
a common place and has been conveniently refuted by Earman
(1995a), among others (see also Nahin 1999 and references
therein). Grandfather-like paradoxes do not imply illogical
situations. In particular, they do not mean that local determinism
does not operate in chronology-violating space-times because it is
always possible to choose a neighborhood of any point of the
manifold such that the equations that represent the laws of
physics have appropriate solutions. Past cannot be changed (the
space-time manifold is unique) but it can be causally affected
from the future, according to GTR. The grandfather paradox, as
pointed out by Earman 1995a, is just a way of showing us that
consistency constraints must exist between the local and the
global order of affairs in space-time. This leads us directly to
the so-called Principle of Self-Consistency (PSC).

\section{Principle of Self-Consistency}

In space-times with CTCs, past and future are no longer globally
distinct. Events on CTCs should causally influence each other
along a time-loop in a self-adjusted, consistent way in order to
occur in the real universe. This has been stated by Friedman et
al. (1990) as a general principle of physics:\\

{\it Principle of Self-Consistency: The only solutions to the laws
of physics that occur locally in the real universe are those which
are globally self-consistent.}\\

When applied to the grandfather paradox, the PSC says that the
grandfather cannot be killed (a local action) because in the far
future this would generate an inconsistency with the global world
line of the time traveler. Just consistent histories can develop
in the universe. An alternative way to formulate the PSC is to
state that (Earman 1995a):\\

{\it The laws of physics are such that any local solution of their
equations that represents a feature of the real universe must be
extendible to a global solution}. \\

The principle is not tautological or merely prescriptive, since it
is clear that local observations can provide information of the
global structure of the world: it is stated that there is a
global-to-local order in the universe in such a way that certain
local actions are ruled out by the global properties of the
space-time manifold.\\

If the PSC is neither a tautological statement nor a
methodological rule, what is then its epistemological status? It
has been suggested that it could be a basic law of physics --in
the same sense that the EFEs are laws of physics-- (Earman 1995a).
This would imply that there is some ``new physics" behind the PSC.
By other hand, Carlini et al. (1995) have proposed, on the basis
of some simple examples, that the PSC could be a consequence of
the Principle of Minimal Action. In this case, no new physics
would be involved (see, however, the objections by Konstantinov
(1996) on this point). Contrary to these opinions, that see in the
PSC a law statement, or at least a consequence of law statements,
we suggest that this principle actually is a {\it metanomological
statement}, like the Principle of General Covariance among others
(see Bunge 1961 for a detailed discussion of metanomological
statements). This means that the reference class of the PSC is not
formed by physical systems, but by laws of physical systems. The
usual laws are restrictions to the state space of physical
systems. Metanomological statements are laws of laws, i.e.
restrictions on the global network of laws that thread the
universe. The requirement of consistency constraints would then be
pointing out the existence of deeper level super-laws, which
enforce the harmony between local and global affairs in
space-time. Just in this sense it is fair to say that ``new
physics" is implied.

\section{Casual loops: Self-existing objects}

\subsection{A bizarre ontology?}

Although the PSC eliminates grandfather-like paradoxes from
chronology-violating space-times, other highly perplexing
situations remain. The most obscure of these situations is the
possibility of an ontology with self-existing objects. Let us
illustrate with an example what we understand by such an object:
\\

{\it Suppose that, in a space-time where CTCs exist, a time
traveler takes a ride on a time machine carrying a book with her.
She goes back to the past, forgets the book in -what will be- her
laboratory, and returns to the future. The book remains then
hidden until the time traveler finds it just before starting her
time trip, carrying the book with her. }\\

It is not hard to see that the primordial origin of the book
remains a mystery. Where does the book come from? This puzzle has
been previously mentioned in philosophical literature by Nerlich
(1981) and MacBeath (1982). Physicists, instead, have not paid
much attention to it, despite the interesting fact that the
described situation is apparently not excluded by the PSC: the
local and global structures of the loop are perfectly harmonious
and there is no casuality violations. There is just a book never
created, never printed, but, somehow, existing in space-time. It
has been suggested (Nerlich 1981) that if CTCs exist, then we are
committed to accept an ontology of self-existing objects: they are
just out there, trapped in space-time. There is no sense in asking
where they are from. Even energy is conserved if we admit that the
system to be considered is not only the present time-slice of the
manifold but rather the two slices connected by the time loop: the
energy removed from the present time ($M_{\rm book}c^2$) is
deposited in the past.\\

It is our view, however, that the acceptance of such a bizarre
ontology proceeds from an incorrect application of the PSC. This
principle is always discussed within the context of GTR, although
actually it should encompass {\it all} physical laws. A fully
correct formulation of the PSC should say {\sl laws of nature}
where in the formulation given above it is said {\sl laws of
physics}. What should be demanded is total consistency and not
only consistency in the solutions of the EFEs. In particular, when
thermodynamics is included in the analysis, the loop of a
self-existing objects becomes inconsistent because, due to
entropic degradation, the final and initial states of the object
do not match. Moreover, even more strange paradoxes, related to
human self-reproduction like the amazing Jocasta paradox (Harrison
1979), can be proven to be non-consistent when the laws of
genetics are taken into account.\\

We shall see that any time loop, described by any physical system
more complex than an ideal fluid in a general wormhole space-time,
undergoes entropy changes that make the loop inconsistent, and the
universe free from self-existing objects.\\

The projection of the derivative of the stress-energy tensor onto
the four-velocity gives, \beq \label{ten2-der}u^i \frac{\partial
T_i^k }{ \partial x^k} = u_i\frac{\partial \tau_i^k }{ \partial
x^k} - T \frac{\partial (\sigma u^i )}{ \partial x^i} + \mu
\frac{\partial \nu^i}{
\partial x^i} =0, \eeq where $\mu=(w-T\sigma)/n$ is called chemical
potential. This equation can be re-written as the total divergence
of a new entropy flux (see Landau and Lifchitz 1975 for details),
\beq \label{ten2-der2} \frac{\partial \sigma^i }{
\partial x^i} =
\frac{\partial } {
\partial x^i}\left(\sigma u^i - \frac \mu T \nu^i\right)
 = -\nu^i\frac{\partial} { \partial x^i}\left( \frac \mu T \right) -
\frac{\tau_i^k}{T} \frac{\partial u^i}{\partial x^k}. \eeq The
explicit forms of $\tau_{ik}$ and $\nu_i$ are given by \beq
\label{tau} \tau_{ik}= -\eta \left( \frac{\partial u_i}{\partial
x^k} + \frac{\partial u_k}{\partial x^i}+ u_k u^l \frac{\partial
u_i}{\partial x^l} + u_i u^l \frac{\partial u_k}{\partial
x^l}\right) - \left( \zeta - \frac 23 \eta \right) \frac{\partial
u^l}{\partial x^l} \left(g_{ik} + u_i u_k \right),\eeq \beq \nu_i=
-\frac \kappa c \left(\frac {nT}{w} \right)^2 \left[ \frac
\partial{\partial x^i} \left( \frac \mu T \right) + u_i u^k \frac
\partial {\partial x^k} \left( \frac \mu T \right) \right].\label{mu}\eeq \\

These two tensors, where $\eta$ and $ \zeta$ are viscosity
coefficients and $\kappa$ is the conductivity, make $\sigma^i$ a
monotonically increasing magnitude, in agreement with the Second
Law of Thermodynamics. The above equations are valid for systems
which are not affected by the gravitational field. In order to
study more general situations, we change the four velocity to
$u^0=1/\sqrt{-g_{00}}$ and $u^\alpha=0$, a system in free fall in
the field, and we also change ordinary derivatives to covariant
ones ($, \;\; \rightarrow \;\; ;$).\\

It is clear that any real fluid will increase its entropy while
traveling along its world line (Landau and Lifchitz 1975, Misner
et al. 1973). There remains the possibility, however, that
particular non-flat space-times have a metric such that a closed
trajectory results in a zero net entropy change due to
cancellation of the two terms in the right member of Eq.
(\ref{ten2-der2}). We will now consider trajectories in general
wormhole space-times in order to explore what happens with
CTCs where real physical systems could be trapped.\\

Many of the analytical wormhole solutions that were presented in
the literature, both for general relativity and other gravity
theories (e.g. Hochberg and Visser 1997; Anchordoqui et al. 1997,
1998), are given in a static, spherically symmetric form, where
the metric is constrained to yield the following interval, \beq
\label{metric} ds^2= - e^{2\alpha} dt^2 + e^{2\beta} dr^2 + r^2
d\Omega^2.\eeq Here, $d\Omega^2=d\theta ^2 + \sin^2 \theta
d\phi^2$ leads to the angular part of $g_{ik}$. Both $\alpha$ and
$\beta$
 are functions (associated with
the metric potentials) only dependent on $r$ because of the
spherical symmetry assumed. These functions have to fulfill
standard wormhole space-time conditions such us the flaring out of
the throat. The specific form of these conditions are unnecessary
for our current purpose. Using the metric (\ref{metric}) one can
immediately prove that \beq \sigma^i_{;\;i}= \frac 1{\sqrt{-g}}
\left( \sqrt{-g}\; \sigma^i
\right)_{,\;i}=\sigma^i_{,\;i}+\sigma^i \frac 1{\sqrt{-g}} \left(
\sqrt{-g} \right)_{,\; i} = \frac \kappa c \left( \frac {n T}{w}
\right)^2\;\; \sum_{i=1}^3 \left( \frac \mu T \right)_{,\;i}^2.
\eeq Here,  sub-indices from 1 to 3 stands for derivatives with
respect to $r,\theta$ and $\phi$. This equation represents the
increase in the entropy flux when a non-ideal fluid is moving in
the gravitational field of equation (\ref{metric}). Any physical
time loop will have non-null length and consequently for the
object in the loop we shall have $\sigma^i_{;\;i}>0$.\\

Note that, due to the form of the metric, all factors of
$\sigma^i_{;\;i}$ involving viscosity coefficients cancel out
themselves (recall that in Eqs. (\ref{tau}) and (\ref{mu}), commas
should be transformed into semi-colons). Only the conductivity
appears in the entropy flux divergence. Note also that
$\sigma^i_{;\;i}$ is positive definite, disregarding the form of
the functions $\alpha$ and $\beta$. There is no way in which a
real fluid, with non-zero conductivity, may traverse the
space-time given in Eq. (\ref{metric}) without increasing its
entropy. Any object using a wormhole as a time machine will be
degraded in a such a way that a CTC can not be formed according to
the PSC. Self-existing objects are, consequently, physically
impossible constructs in wormhole space-times .\\

\section{Casual loops: Auto-generated information}

\subsection{Information coming from nothing?}

Let us now consider again the time traveler mentioned before, and
suppose that when she arrives to the past she looks for the author
of the book she has brought with her. Imagine that once the author
is found, she gives the book to him before he actually wrote it,
and then comes back to the future. The book stays in time
traveler's past and naturally ages as time goes by. The yet-to-be
author reads and likes the book, and decides that the material
deserves wide publicity. So, he types the text and sends the
manuscript to the editor signed with his name. The editor agrees
with the publication and the book is edited. Some years latter the
time traveler takes one copy and carries it to the past, where she
gives it to the author. Where does the information contained in
the book come from?. Here, there is no problem with physical time
loops, since the object (the book) normally ages
with time. The PSC seems not to be violated in any respect. \\

This kind of casual loops are called information paradoxes (see,
for instance, Visser 1996 and Nahin 1999). When a space-time is
such that it admits chronology violating regions there seems that
information can be generated from nothing. Lewis (1976) has
already written on casual loops involving just information
transfer. He concluded that there is simply no answer to the
question on the origin of the information, and that casual loops
of this kind are not too different from other situations we are
confronted with, such as the origin of the universe or the decay
of a tritium atom. A similar view is sustained by Levin (1980),
who compares casual loop paradoxes to questions about where
anything originally comes from. CTCs can confront us with a way of
producing knowledge that conflicts with the principles of
philosophy of science, since knowledge can come into existence
without any evolutionary process (Deutsch 1991). However, it is
important to notice that information paradoxes are not paradoxes
at all, in the sense that they do not imply any contradiction with
logical or even physical laws. They just seem to contradict our
common sense, but this is
not enough to rule them out. \\

In an recent twist of the problem, Smith (1997) argues that a
comparable example of information coming from nothing could be
described as follows. Suppose that someone says something trivial,
but other person mishears him, thinking that he has said something
profound. Smith asks, ``where does this profound idea come from?".
It seems to have been ``spontaneously" generated and, we are told,
this would be similar to what happens with information in a CTC.
We sustain, on the contrary, that both situations are completely
different. The formation of ideas and concepts is a brain process.
We say that an idea has emerged when certain neural plastic
networks are activated. The activation mechanism, although not
clear, is surely far from linear. Small changes in the input
signals can trigger quite different responses. In particular, a
small perturbation in the process of hearing a speech can result
in a different chain of thoughts in the listener (see Bunge 1980
and references therein for more details on the ideation process).
It is clear then that in Smith's situation, the idea comes from
electro-chemical processes in a brain and not from nowhere. Even
if we consider the information ``carried" by the idea, this
information does not come from nowhere; it is produced in the
brain of a human being. Furthermore, Smith's method of producing
information is very limited: it can not be used to transmit an
information volume comparable to a whole book, but just some
phrases, probably ultimately related to the course of a previous
conversation. The phenomenon quoted by Smith is, besides, local in
time, whereas information produced by CTCs is not. In casual
loops, information simply exists as a consequence of the peculiar
topology of the space-time manifold.

\subsection{Information loops and the PSC}

Consider the local light cone of a time traveler. There are three,
and only three, possible final destinations for a time trip. No
matter the way in which the travel is done, the final destiny of
the traveler can only be one out of three possibilities: either
{\bf a} she arrives inside her past light cone, or {\bf b} she
arrives on the light cone, i.e. at a distance equal (in
geometrized units) to the time that light would take to come back
to the initial position through normal space, or {\bf c} she
arrives at a point out of her initial light cone. For the first
two possibilities, an information loop is obviously unavoidable.
For the third one, when the traveler arrives at a point out of her
initial past light cone, there is no information loop with only
one time trip. Traveling at the speed of light, information sent
back to the initial position of the time machine will always
arrive after the time trip commenced. But even in the last case,
if more than one time machine are available (for instance as in
the situation known as a Roman ring (Visser 1997), where there are
two wormholes in relative motion) the past might be affected by
information flux from the future. However, with just one time
machine, affecting the past is impossible in case {\bf c}.
Advocates of chronology protection could then invoke that just
this kind of time travel is possible in nature. This requirement
could be stated as a {\it Conjecture of Past Shielding:} every
single-time-machine time trip would be such that the two extreme
points are always separated by a space-like interval. In order to
enforce this conjecture it could be pointed out that, if natural
wormholes exist at all, their number would be so scarce that a
Roman ring configuration (Visser 1997) would well be practically
impossible (Torres 1998a, b; Anchordoqui 1999, see -on the same
topic- Eiroa et al. 2000 and
Safonova et al. 2000a, b).\\

However, the conclusion is that whatever the final destiny of the
time-traveler is, in principle, she could always affect her own
past. Otherwise stated, chronology-violating space-times generally
admit information loops: they cannot be excluded on the only basis
of the PSC.\\

Although the PSC do not preclude that within chronology-violating
regions information steaming from the future can affect the past,
it at least imposes constraints on the way this can be done. In
fact, any physical process causally triggered by the backwards
information flux must be consistent with the past history of the
universe. This means that if a time traveler goes back to his past
and tries, for instance, to communicate the contents of the theory
of special relativity to the scientific community before 1905, she
will fail {\it because} at her departure it was historically clear
that the first paper on special relativity was published by Albert
Einstein in June 1905. The details of her failure will depend on
the details of her travel and attempt, in the same way that the
details of the failure of a perpetual motion machine depends on
the approach used by the imprudent inventor. All we can say {\it a
priori} is that the laws of physics are such that these attempts
{\it cannot} succeed and information cannot propagate arbitrarily
in space-time. It is precisely due to the PSC that we know that
our past is not significantly affected from the future: we know
that till now, knowledge has been generated by evolutionary
processes, i.e. there is small room in history for information
loops. This does not necessarily mean that the same is valid for
the entire space-time.

\section{Conclusions}

The field equations of GTR, as well as those of other classical
theories of gravity, have solutions that allow the formation of
closed time-like curves. Some of these solutions occur under
conditions that could happen in the real universe, at least as far
as our present knowledge let us see. Although a full theory of
quantum gravity could, perhaps, provide some effective mechanism
to enforce chronology protection, the success of the classical
theory over a very wide range of applications makes commendable
the study of the expected behaviour of physical systems in the
presence of CTCs. As remarked by Thorne (1992), through pushing
the theory to its most extreme predictions, we can get important
insights of its limitations and, hopefully, the ways to supersede
them. From this point of view, time travel is more than a mere
justification for theoretical speculation; it is a conceptual tool
that can be used to probe the deepest levels of General Relativity
and the picture of the world that emerges from this theory; it is
an epistemological instrument that can lead us to an important
clarification of the very foundations of space-time theories and
their implications.\\

In this paper we have shown that the Principle of Self-Consistency
excludes the possibility of a bizarre ontology with self-existing
objects. We also have suggested that this principle is a
metanomological statement that enforces the global-to-local
harmony in the network of laws that determine the evolution of
physical systems in space-time. The information flux from future
to the past is possible in GTR, although restricted by consistency
constraints. This means that past and future are not longer
globally distinct when CTCs occur and that causality can globally
-although never locally- operate backwards.

\section*{Acknowledgments}
This work has been supported by the agencies CONICET and ANPCT
(PICT 03-04881) as well as by funds granted by Fundaci\'{o}n
Antorchas to both authors.


\end{document}